\documentclass[12pt]{iopart}

\usepackage{graphicx}

\begin{document}

\jl{7}

\title[Independent neurons representing a finite set of stimuli]
{Independent neurons representing a finite set of stimuli:
dependence of the mutual information on the number of units sampled}

\author{In\'es Samengo}

\address{International School of Advanced Studies\\ Via Beirut 2 - 4,
(34014) Trieste, Italy}

\begin{abstract}
We study the capacity with which a system of independent neuron-like
units represents a given set of stimuli. We assume that each neuron
provides a fixed amount of information, and that the information
provided by different neurons has a random overlap. We derive
analytically the dependence of the mutual information between the set
of stimuli and the neural responses on the number of units sampled.
For a large set of stimuli, the mutual information rises linearly
with the number of neurons, and later saturates exponentially at its
maximum value.
\end{abstract}


\submitted

\maketitle


\section{Experimental measures of redundancy}

The capacity with which a population of neurons, or neuron-like
units, can represent a set of elements in the outside world is an
important issue both for experimental studies of neural coding and
for theoretical analysis of neural network models. In the former
situation, typically a discrete set of $p$ stimuli is presented to
a subject while the activity of a population of $N$ neurons is
recorded. The measured response can be
represented as an $N$ dimensional vector ${\bf r}$, whose components stand
for the activity of individual
neurons, computed over a predefined time window.
This activity is expected to be selective, at least to some
degree, to each one of the stimuli. The degree of selectivity can
be quantified by the mutual information between
the set of stimuli and the responses (Shannon 1948)
\begin{equation}
I = \sum_{s = 1}^p P(s) \sum_{{\bf r}} P({\bf r} | s) \ \log_2
\left[ \frac{P({\bf r} | s)}{P({\bf r})} \right],
\label{inf}
\end{equation}
where $P(s)$ is the probability of showing stimulus $s$,  $P(
{\bf r} | s)$ is the conditional probability of observing response
${\bf r}$ when the stimulus $s$ is presented and
\begin{equation}
P({\bf r}) = \sum_{s = 1}^p P(s) \ P({\bf r} | s).
\end{equation}
The mutual information $I$ characterizes the mapping between the $p$
stimuli and the response space, and represents the amount of information
conveyed by ${\bf r}$ about which of the $p$ elements was shown.

If each of the stimuli evokes a unique set of responses, i.e. the responses
are different for different stimuli, then Eq. (\ref{inf}) reduces
to the entropy of the set of stimuli $H\{s\} = - \sum_s P(s) \log_2 P(s)$.
When the stimuli are equiprobable, the subject is exposed to the most
variable presentation of the $p$ elements, and the entropy of the stimuli
reaches its maximum value, that is, $\log_2(p)$. For other choices of
$P(s)$, the mutual information is still bounded by $H\{s\}$, which in
turn, is less than $\log_2 p$.

On the other hand, if a response ${\bf r}$ may be evoked by more than
one stimulus, the mutual information is less than $H\{s\}$.
In the extreme case where the responses are independent of the stimuli,
$I = 0$. In short, the mutual information quantifies how precisely can
any single stimulus be identified out of the $p$ possible ones, by
measuring the response of the $N$ units.

Here, we are interested in describing the dependence of the mutual
information with the number of neurons sampled. Such a dependence is
crucial in quantifying the redundancy between the messages carried
by different cells. Since information is an additive quantity,
if different neurons provide independent information, we expect
$I(N)$ to grow linearly with $N$. In fact, any departure of linearity
is a sign of either a redundant or a synergetic coding (depending
on whether the behaviour is sub or supra-linear). However,
before drawing conclusions about the way information is shared
among the neurons, it is important to understand to what extent
the experimental design, more precisely, the set of stimuli
chosen, determines by itself the shape of $I(N)$. As stated above,
the maximum information that can be extracted from the neural
responses is $H\{s\}$. It is clear that if we have a set of
neurons that already provides an information very near to this
maximum, by adding one more unit we will gain no more than
redundant information. In other words, we have reached a regime
where the neural responses correctly identify all of the stimuli.
But we cannot deduce from this that the
representational capacity of the responses remains unchanged
when the number of neurons increases. One should rather realize
that the task itself is no longer appropriate to test the way
additional neurons contribute in the encoding of the stimuli.

\begin{figure}
\begin{center}
\scalebox{.7}{\includegraphics{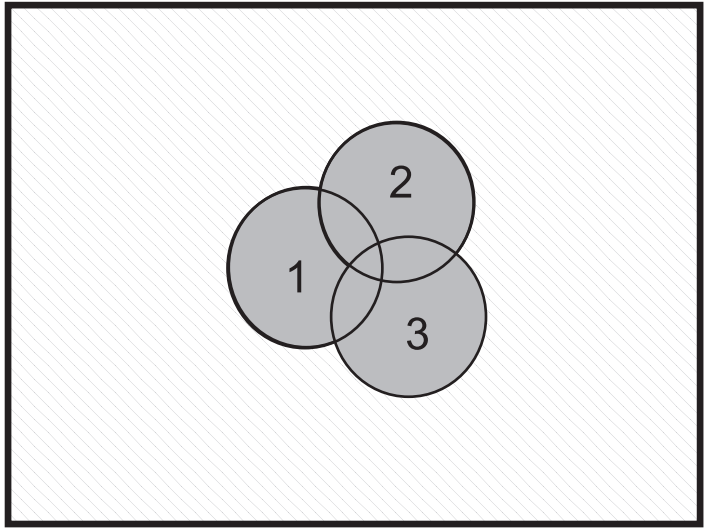}}
\end{center}
\caption{Schematic representation of Gawne and Richmond's process
(1993).
The area of each circle represents the amount of information
$I(1)$ provided by a single neuron. Different circles correspond to
different neurons. Any two pair of neurons share an overlap
$y I(1)$.}
\label{f1}
\end{figure}

Gawne and Richmond (1993) have considered this issue quantitatively,
when recording from pairs of neurons. They presented a simple model which
yields an analytical expression for $I(N)$ under the assumption that
each neuron provides a fixed amount of information $I(1)$, and that a fixed
fraction $y$ of such an amount is redundant with the information conveyed by
any other neuron. In figure \ref{f1}
we reproduce their scheme, where all pairs
of neurons are supposed to share the same redundancy $y I(1)$.
Their model yields
\begin{equation}
I(N) = \frac{I(1)}{y}\left[1 - (1 - y)^N \right].
\label{gr}
\end{equation}
Here, $I(1)$ and $y$ are considered independent quantities.
Gawne and Richmond (1993) measured both of them, using a single
electrode to record the activity of pairs of nearby neurons.
They got an average value for $I(1) = 0.23$ bits, and their mean
redundancy $y = [2 I(1) - I(2)]/I(1) =  0.2$. According to equation
(\ref{gr}), as $N$ increases, $I(N)$ approaches a limiting value
equal to $I(1) / y$. In their formulation, no attention is payed
to the fact that $I(\infty)$ should approach $H\{s\}$, at least if
the subject is behaviourally able to identify every single stimulus.
According to the measured values of $y$ and $I(1)$, Gawne and
Richmond calculated $I(\infty) = 1.15$. Since in their case
$H\{s\} = \log_2 p = 5$, they concluded that the assumption that all
pairs of neurons share the same amount of redundant information,
as measured by nearby neurons,  was wrong.

What appeared important, then, was to go beyond what could be
extrapolated from the shared information between pairs of cells,
and measure directly the information that could be extracted from
large populations. From the experimental point of view, this meant
to measure the activity of a large number of units, ideally, with
simultaneous recordings. Rolls \etal (1997) have performed the
experiment measuring the responses from cells in the inferior
temporal cortex of a macaque, when exposed to $p$ visual stimuli. They
recorded one neuron at a time, and therefore, were not able to capture
the correlations in the neural responses present in every single trial.
One should bear in mind,
however, that such non-simultaneous recordings correctly describe the
information carried by the firing rates of the neurons, and are
exact in the limit of short time windows (Panzeri \etal, 1999a).

\begin{figure}
\begin{center}
\scalebox{.7}{\includegraphics{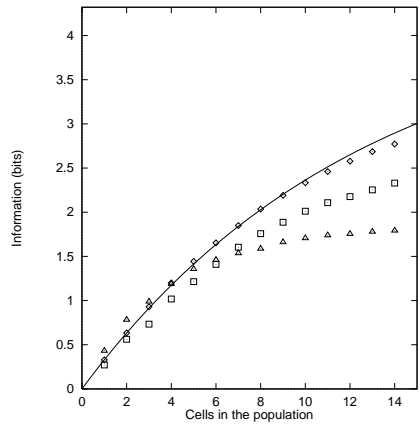}}
\end{center}
\caption{
Mutual information extracted from neural responses from the inferior
temporal cortex of a macaque when exposed to $p$ visual stimuli
(from Rolls \etal 1997). Diamonds correspond to $p = 20$, squares
to $p = 9$ and triangles to $p = 4$. The graph is
plotted as a function of the neurons considered. We see that initially
the information grows linearly, and eventually slows down For $p = 4$,
the curves clearly saturate at $\log_2 p = 2$.
}
\label{f2}
\end{figure}

In figure \ref{f2}
we show their estimation of $I(N)$, after a decoding procedure.
In their experiment, stimuli were equiprobable, so $H\{s\} = \log_2 p$.
Diamonds correspond to $p = 20$, squares to $p = 9$ and triangles
to $p = 4$. The graph is plotted as a function of the number of
neurons sampled. Each line is an average of all the possible curves
that could be made, depending on the order in which the neurons are
selected. We see that initially the information grows linearly, and
later descelerates. For $p = 4$ there is a clear saturation at
$\log_2 4 = 2$ bits. For $p = 9$ and $p = 20$ the behaviour of the curves
is compatible with the predicted asimptotes, at $\log_2 9 \approx 3,16$ bits
and $\log_2 20 \approx 4,23$ bits. This experiment shows that the pool of
neurons sampled appears to reach the entropy of the set of stimuli, showing
that the activity of a sufficiently large number of cells carries the
information needed  to correctly identify every single element.

In order to explain their results, Rolls \etal have considered a more
constrained model than the one of Gawne and Richmond, that
further assumes that $y = I(1) / \log_2 p$. This means that the
only parameter to be measured is now $I(1)$.
Their approach also leads to equation (\ref{gr}), but now
$I(\infty) = \log_2 p$.
A fit of their expression is shown by the full line in
figure \ref{f2}, for the case of 20 stimuli.
Although their choice for the overlap $y$ might seem arbitrary,
it is the only one that leads to the correct asymptotic behaviour.
In this paper we show that such a choice is also the average
redundancy if the information provided by a pair of neurons has a
random overlap.

In the following section we set the problem in a probabilistic
framework, namely we calculate the probability distribution that $N$
neurons provide an amount $I$ of information. We also show that equation
(\ref{gr}) is the mean value of such a distribution, and we obtain
its standard deviation. Next, in section 3 we extend our model to a
more complex case, and we end in section 4 with some concluding
remarks.


\section{A phenomenological probabilistic approach}

Our aim is to derive the probability $f(N, I)$ that $N$ neurons
provide an amount $I$ of information when firing in response
to a set of $p$ stimuli. We do so in terms of a single parameter,
$I(1)$ defined as the mean single-neuron information.

\begin{figure}
\begin{center}
\scalebox{.7}{\includegraphics{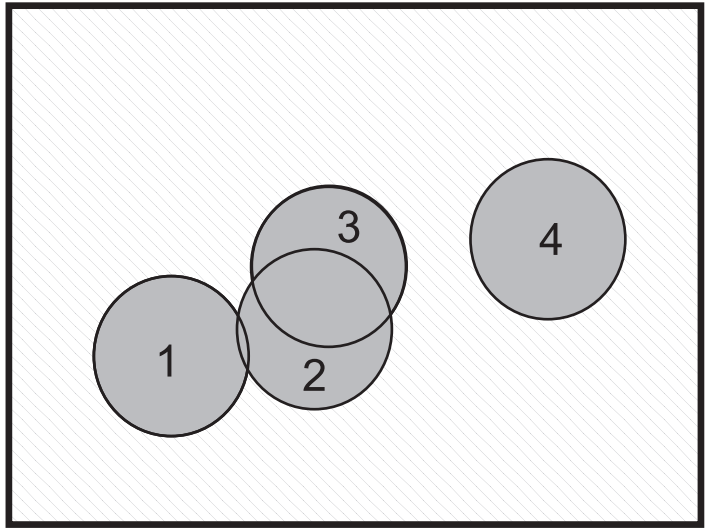}}
\end{center}
\caption{Schematic representation of the way the information
provided by different neurons overlap with one another. The
stripped area represents the maximum information, namely,
$H\{s\}$. Each neuron provides an information $I(1)$, that
may fall anywhere in the rectangle. In the case shown here,
a significant fraction of the information provided by neurons
2 and 3 is redundant.}
\label{f3}
\end{figure}

We assume that the information extracted from a given
neuron has a uniform probability of representing any portion of the
maximum information. Figure \ref{f3} shows a pictorial representation
similar to the one introduced by Gawne and Richmond, with the difference
that now the information provided by
every single cell may fall anywhere in the striped area, and with
a uniform probability distribution.

For the moment, we assume that the information provided by neuron
$N + 1$ is either entirely redundant with the one carried by the
previous $N$, or the whole of it is new. In other words,
neurons provide information in indivisible blocks of size $I(1)$.
Later we allow for the possiblility of partial overlap.
At the present stage, however,  $I$ can only increase in
steps of a fixed size, namely of $I(1)$.

With these assumptions, the probability that the
information extracted from the response of the neuron $N + 1$ is
redundant with the information provided by neurons
$\left\{1, 2, ... N\right\}$ is equal
to the ratio $I(N) / H\{s\}$. Accordingly, the probability that neuron
$N + 1$ gives new information is $[H\{s\} - I(N)] / H\{s\}$.

Therefore, the probability that $N + 1$ neurons provide an information
$I + I(1)$ has two different contributions. On one hand, it may happen
that the first $N$ neurons had already provided an information
$I + I(1)$, and the information of neuron $N + 1$ comes to be redundant
to the previous information. Alternatively, it may be that the response
of the first $N$ neurons gave an information equal to $I$, and the new
neuron provides the extra $I(1)$. We therefore
write a recurrence equation for the distribution $f$,
\begin{equation}
f(N + 1, I + I(1)) = \frac{I + I(1)}{H\{s\}} f(N, I + I(1)) +
\frac{H\{s\} - I}{H\{s\}} f(N, I).
\label{recurr}
\end{equation}
As a first step, and in order
to make the notation simpler, we choose to measure the information
in units of $I(1)$. In these units, the maximum information reads
$I_\infty = H\{s\} / I(1)$. Thus, Eq. (\ref{recurr}) reads
\begin{equation}
f(N + 1, I + 1) = \frac{I + 1}{I_\infty} f(N, I + 1) +
\frac{I_\infty - I}{I_\infty} f(N, I).
\label{recu1}
\end{equation}
This recurrence equation has to be solved with the initial
conditions
\begin{eqnarray}
f(0, I) &=& \delta_{0, I}, \nonumber \\
f(N, 0) &=& \delta_{0, N},
\label{borde}
\end{eqnarray}
where $\delta_{a,b} = 1$ if $a = b$, and $\delta_{a, b} = 0$
otherwise. The first condition states that no neurons can only
give no information, while the second indicates that if there is
no information, then, for sure, there are no neurons.

We define a generating function
\begin{equation}
G(x, y) = \sum_{N = 0}^{\infty} \sum_{I = 0}^{\infty} f(N, I) x^N
y^I.
\end{equation}
Clearly,
\begin{equation}
f(N, I) = \frac{1}{N! \ I!} \ \left. \frac{\partial^{N + I} \ G}
{\partial^N x \ \ \partial^I y}\right|_{x = y = 0}.
\end{equation}
Therefore, if equation (\ref{recurr}) is transformed into an
equation for $G$, and this last equation is solved, the
probability $f$ can be readily calculated, by derivation.
Moreover, if we define the mean values
\begin{equation}
\langle I^j \rangle = \sum_{I = 0}^\infty I^j f(N, I),
\end{equation}
we observe that the first two moments can be obtained with
the equalities
\begin{eqnarray}
\left.\frac{\partial G}{\partial y}\right|_{y = 1} &=&
\sum_{N = 0}^{\infty} \langle I(N) \rangle \ x^N \nonumber \\
\left.\frac{\partial^2 G}{\partial y^2}\right|_{y = 1} &=&
\sum_{N = 0}^{\infty} \left[ \langle I^2(N) \rangle -
\langle I(N) \rangle \right] \ x^N,
\label{medios}
\end{eqnarray}
namely, making a series expansion of derivatives of $G$.

It can be shown that the recurrence relation (\ref{recu1}) with its
initial conditions (\ref{borde}) are equivalent to a
differential equation for $G$,
\begin{equation}
\frac{x y}{I_\infty}(1 - y)\frac{\partial G}{\partial y} +
(x y - 1) G + 1
= 0 ,
\label{part}
\end{equation}
to be solved with the border conditions
\begin{equation}
G(0, y) = G(x, 0) = 1.
\label{part1}
\end{equation}
Morover, since $f$ is a normalized probability distribution,
\begin{equation}
G(x, 1) = \sum_{N = 0}^\infty x^N = \frac{1}{1 - x}.
\label{norm}
\end{equation}
The condition $f(N, I > I_\infty) = 0$
implies that $G$ is a polynomial in $y$.

The differential equation (\ref{part}) for $G$ does not involve
derivatives in $x$. Therefore, for each value of $x$ one has an
ordinary first order differential equation in $y$, which can be readily
integrated. The result is
\begin{eqnarray}
G(x, y) &=& \frac{y}{1 - x} \ _2F_1\left[1, 1 - I_\infty; 1 + I_\infty
\frac{1 - x}{x}; 1 - y \right]  \nonumber \\
& & +\frac{I_\infty(1 - y)}{x + I_\infty(1 - x)} \ _2F_1 \left[1, 1 - I_\infty;
2 + I_\infty \left(\frac{1 - x}{x} \right); 1 - y \right],
\label{G}
\end{eqnarray}
where $_2F_1$ is Gauss' Hypergeometric Function (Abramowitz, 1972).
It can be easily seen that $G$ fulfills both the border conditions
in (\ref{part1}) and the normalization constraint (\ref{norm}).
Morover, if $I_\infty$ is a positive integer, then the hypergeometric
functions in (\ref{G}) become polynomials in $y$. It may seem, in
a first thought, that the requirement of $I_\infty$ being an integer
is too restrictive. However, if the whole of the striped rectangle
in figure \ref{f3} is supposed to be filled by blocks of area $I(1)$, it
is in fact necessary to have a total area ($H\{s\}$) that is an
integer multiple of $I(1)$.

By simple derivation, we get
\begin{equation}
\left. \frac{\partial G}{\partial y} \right|_{y = 1} =
\frac{x}{(1 - x) \left[1 - x(I_\infty -1)/I_\infty \right]}.
\end{equation}
This expression involves a linear term in $x$ and two factors
that can be written as geometric series. It is therefore easily
expanded in powers of $x$. Using equation (\ref{medios})
we obtain
\begin{equation}
\langle I(N) \rangle = \sum_{j = 0}^{N - 1} \left(\frac{I_\infty - 1}
{I_\infty} \right)^j = I_\infty \left[1 - \left(\frac{I_\infty - 1}{I_\infty}
\right)^N \right].
\label{huk}
\end{equation}
If instead of writing $I$ in units of $I(1)$ we turn back to
bits, equation (\ref{huk}) coincides with (\ref{gr}). In order
to state the equivalence of the two, we have to set
\begin{equation}
y = I(1) / H\{s\} =  1 / I_\infty,
\label{conecta}
\end{equation}
as was done by Rolls \etal (1997).

\begin{figure}
\begin{center}
\scalebox{.5}{\includegraphics{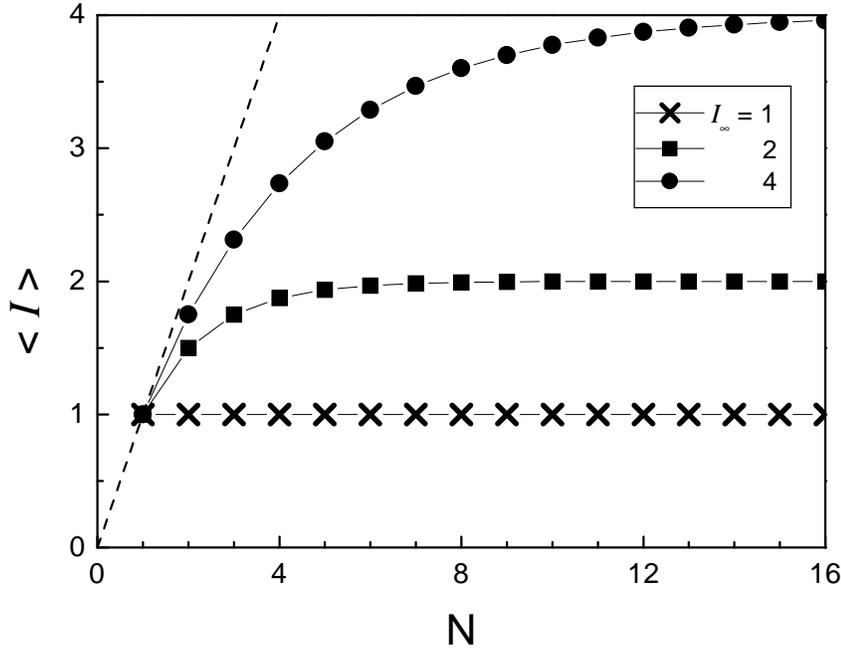}}
\end{center}
\caption{Mean information as a function of the number of neurons
sampled, for different values of $I_\infty = H\{s\} / I(1)$. The
crosses $(\times)$ represent the trivial case of $I_\infty = 1$.
As $I_\infty$ grows, the curves rise. In the limit of $I_\infty \to
\infty$, they approach the dashed line. In all cases, $\langle
I (N = 0)\rangle = 0, \langle I (N = 1) \rangle = 1$ and
$\langle I (N \to \infty) \rangle = I_\infty$.}
\label{f4}
\end{figure}

In figure \ref{f4}
we show $\langle I \rangle$ as a function of
$N$, for different values of
$I_\infty$. The crosses represent the case of $I_\infty = 1$,
where a single neuron provides the maximum
information. Obviously, additional cells produce no
variation in the mutual information.
The squares and the circles represent
$I_\infty = 2$ and $I_\infty = 4$, respectively. The dashed line shows
the behaviour in the limit of $I_\infty \to \infty$. All the curves
saturate at $I_\infty$.

Equation (\ref{huk}) implies that when $I_\infty \gg 1$
(or equivalently $I(1) \ll H\{s\}$), the mutual information rises almost
linearly with $N$. In fact, expressing $I(N)$ in bits,
\begin{equation}
\lim_{H\{s\} \gg I(1)} I(N) = N I(1).
\end{equation}
Therefore, we conclude that for $I(1) \ll H\{s\}$
initially different neurons provide independent information.
In the schematic view of figure \ref{f3}, when the number of neurons is small,
they will probably occupy different areas of the striped region.
The slope of the initial rise is a measure
of the mean information provided by a single unit.
The deviation from a strictly linear behaviour is given by
the mean pairwise overlap $y = [2 I(1) - I(2)] /I(1)$.
In all cases, as the number of units is increased,
almost the whole of the rectangle is covered, and therefore,
the mean information approaches its ceiling.
In fact, as $I(1)/H\{s\}$ increases, the initial linear
rise is no longer observed, as shown by the squares and in the limit
case of the crosses of figure \ref{f4}.

We  now turn to the second moment of the distribution $f$. We
define the dispersion
\begin{equation}
\sigma^2(N) = \langle I^2(N) \rangle - \langle I(N) \rangle^2.
\label{sigcucu}
\end{equation}
Thus, $\sigma$ quantifies how much variability one expects to find in
different trials of the random process depicted in figure \ref{f3}.
Or, from the point of view of the experimentalist, the changes that may
be expected when $I(N)$ is measured using different sets of
$p$ stimuli---keeping them all within the broad class of
stimuli to which the cells are supposed do be responsive.
Making use of equation (\ref{medios}), we get
\begin{eqnarray}
\sigma^2(N) &=& I_\infty \left(I_\infty - 1 \right)
\left\{1 - 2 \left(\frac{I_\infty - 1}{I_\infty} \right)^N +
\left(\frac{I_\infty - 2}{I_\infty} \right)^N \right. - \\
& & \left. -
\left[1 - \left(\frac{I_\infty - 1}{I_\infty} \right)^N \right]
\left[1 - \left(\frac{I_\infty - 1}{I_\infty}\right)^{N - 1} \right]
\right\},
\nonumber
\end{eqnarray}
where, for simplicity, $\sigma^2$ is written in units of
$[I(1)]^2$. As expected, $\sigma(N)$ vanishes for $N = 0, 1,
\infty$. In figure
\begin{figure}
\begin{center}
\scalebox{.5}{\includegraphics{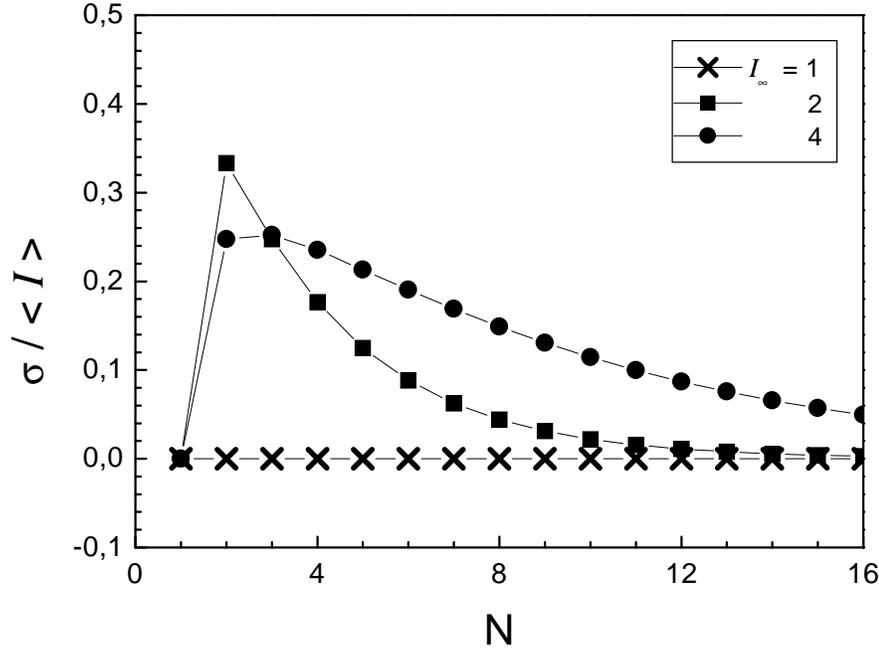}}
\end{center}
\caption{Relative dispersion $\sigma / \langle I \rangle$
(see equation (\ref{sigcucu}))as a function of the number of
neurons sampled, for different values of $I_\infty = H\{s\} / I(1)$.
When $I_\infty = 1$ and $I_\infty \to \infty$, the dispersion vanishes.
For intermediate values, we observe a rise in the standard
deviation. For all values of $I_\infty$ we have that $\sigma (N = 0) =
\sigma(N = 1) = \sigma(N \to \infty) = 0$.}
\label{f5}
\end{figure}
\ref{f5} we plot the relative dispersion $\sigma / \langle I \rangle$
as a function of $N$. We observe that for $I_\infty = 1$ the standard
dispersion vanishes, for all $N$. As $I_\infty$ grows, $\sigma /
\langle I \rangle$ rises. Its maximum value is reached for $N_{{\rm max}}
\approx I_\infty$. For very large $I_\infty$, the curve flatterns again.


\section{Allowing for partial overlaps}

Let us now re-formulate the model to allow for the possibility of
partial overlap between the information conveyed by any two neurons.
Here we present one possible way to do so. Instead of thinking that when
one more neuron is considered a whole block of information $I(1)$ is
added, we imagine that the same amout $I(1)$ appears in $M$ pieces
of size $I(1) / M$. In other words, the information provided by
$N$ neurons is the result of sampling at random the whole of the available
information a number $N' = M N$ of times, each sample of size
$I(1)/M$. Ultimately, we shall put $M \to \infty$.
We coin $g(N', I)$ for the probability distribution that after $N'$
steps, an amount $I$ of information is extracted. Clearly, and by
analogy with equation (\ref{recurr}), $g$ obeys the recurrence relation
\begin{equation}
\fl
g(N' + 1, I + I(1)/M) = \frac{I + I(1)/M}{H\{s\}}
g(N', I + I(1) / M) + \frac{H\{s\} - I}{H\{s\}} g(N', I).
\label{recu2}
\end{equation}
Since we are interested in the information provided by $N$
neurons---and not by $N'$ steps---we relate the two probabilities
stating
\begin{equation}
f(N, I) = g(M N, I).
\label{junta}
\end{equation}
This subdivision of the process has two main
consequences. First, and as desired, the overlap between the information
conveyed by any two neurons in a particular realization is no longer either
zero or $I(1)$. Now, the overlap can be any $k I(1) / M$, for
integer $k$ varying in $[0, M]$. But in addition, now single neurons do
not provide a fixed amount of information. The $M$ blocks of information
conveyed by a single neuron may overlap with one another. Therefore,
the information provided by just one neuron obeys a distribution ranging
from $I(1)/M$ (when all the blocks overlap) up to $I(1)$ (when there
is no overlap at all).

In order to solve (\ref{recu2}) we choose to measure the information
in units of $I(1)/M$. We define
\begin{eqnarray}
I' &=& \frac{I}{I(1) / M} \nonumber \\
I_\infty' &=& \frac{H\{s\}}{I(1) / M}.
\end{eqnarray}
If the recurrence relation for $g$ is written as a function of $N'$ and $I'$,
equation (\ref{recu1}) is obtained, with the only difference that now all
variables appear primed. This means that the whole procedure of the
previous section can be applied. Using equation (\ref{junta}) to go back
to $f$ and taking the limit $M \to \infty$, we get
\begin{equation}
\lim_{M \to \infty} \langle I \rangle =
I_\infty \left(1 - e^{- N / I_\infty}\right).
\label{blip}
\end{equation}
As stated above, in this case, $\langle I(N = 1) \rangle$ is not necesarily
equal to $I(1)$. Moreover, the whole curve $\langle I(N)\rangle$ differs,
in general, from (\ref{huk}). More precisely, the factor
$[(I_\infty - 1) / I_\infty]^N$
has now been replaced by $\exp(- N / I_\infty)$.
We observe, however, that as $I_\infty$ increases,
the two results coincide.
It  can be also shown that in the limit $M \to 0$ the
dispersion $\sigma$ vanishes.

The procedure we have devised in order to include the possibility
of partial overlaps is equivalent to a re-scaling of
the variables $N$ and $I(1)/H\{s\}$. It may be seen that as
$M$ grows, the distribution $f(N, I)$ becomes more and more
concentrated in its mean value $\langle I(N) \rangle$. In fact,
had we attempted to solve equation (\ref{recurr}) by taking the
continuous limit in the size of the increments of both
$N$ and $I$, we would have obtained a solution that
is a delta function vanishing for all $I$ and $N$, except
for those that obey the equation $I = I_\infty [1 - \exp(-N/I_\infty)]$,
which is exactly the mean value predicted by (\ref{blip}).


\section{Implications for experimental measurements}

We have calculated the probability
distribution that $N$ neuron provide an amount $I$ of information.
We have seen that the mean value of such a distribution is given by
equation (\ref{gr}) already proposed by Gawne and Richmond (1993), and
further constrained by Rolls \etal (1997).
The remarkably good accuracy with which equation (\ref{gr})
can fit the real data (see figure \ref{f2}) suggests that
the present approach, although phenomenological,
captures the relevant aspects of the way the information about the
identity of the stimulus shown is shared among neurons.

The probability distribution characterizes an ensamble of equivalent
processes, all of which fulfill two conditions. In the first place,
every unit is supposed to provide a fixed amount of information
$I(1)$. Rigourously speaking, this assumption is not true, since there is no
doubt that some neurons are more informative than others. Our main interest,
however, is to derive how the information scales with the
number of neurons. We aim at predicting the type of behaviour shown
in figure \ref{f2}. In order to obtain a somewhat universal trend
that does not depend on the particular identity and order in which
neurons are chosen, it is useful to work with the average single-unit
information. Also in the plot of figure
\ref{f2}, each point represents an average over all
the possible selections of $N$ neurons ($N$ ranging from one to fourteen),
out of the fourteen sampled ones.

Of course, the suposition that every unit provides
$I(1)$ bits of information does not mean that there is no trial to
trial variability. Whatever the size of the variability---in other
words, whatever the distribution $P(r_j|s)$---one calculates,
using equation (\ref{inf}), the information provided by unit $j$. By
averaging this quantity on all $j$, $I(1)$ is obtained.

The second main assumption is that any one neuron provides a random
portion of the total available information. When many trials of
such a stochastic process are averaged together, one deduces that
any two neurons share, on average, a fixed fraction
$y$ of $I(1)$. Moreover, when a third neuron is added, the
portion of the information that is shared between the three of them
is the same fraction $y$ of the mean redundancy between two neurons.
In principle, the two fractions need not be the same. In this sense,
the present approach is the simplest choice that could be made.

What does this random overlap hipothesis mean, from the point of view
of neural operation? If two neurons are strongly correlated
(for expample, if they share an appreciable fraction of their imputs)
they often provide the same information---more often than by chance.
If, on the contrary, the set of neurons is precisely designed
so that each unit represents
a different feature of the stimuli then, if the stimuli evenly cover the
feature space, the information provided by different neurons will seldom
overlap---this means, less frequently than by chance.
The random overlap hipothesis is, in some way,
a null hipothesis. No special organization is assumed,
neither in making the neurons particularly redundant nor synergetic.
This phenomenological level of description should be contrasted to
others approaches, where the correlations among units
are described in more detail (Oram \etal 1998, Abott and Dayan 1999,
Panzeri \etal 1999b, Karbowski 2000).
As limited as it might seem, our approach has the appealing property
of being analyticaly tractable. Moreover, the resulting $I(N)$
depends on a single
parameter $I_\infty$.

An alternative to the assumptions made in the present paper would be
to model explicitely the conditional probabilities $P({\bf r}|s)$ and
calculate $I(N)$ using equation (\ref{inf}), as done in Samengo and
Treves (2000). This is a much more detailed level of description, and
allows the exploration of different coding strateggies, for example,
a localized scheme (sometimes called a grandmother-cell encoding) shows
a particular behaviour of $I(N)$, that differs from the one observed in
a distributed code. However, such an approach is also much more arbitrary,
since a precise model of the unknown parameters shaping the neural responses
is needed.

The formalism presented here predicts an average redundancy
$y = I(1) / H\{s\}$
entirely determined by ceiling effects. In other words, if the
average information provided by each neuron is $I(1)$, then just
because the maximum information is finite, there will be a pairwise
overlap of $y I(1)$. Therefore, if in an actual experiment the mean
redundancy measured is similar to $I(1)/H\{s\}$, this means that
there is not much redundancy or synergy in the coding scheme.
Such a value stems from the null hypothesis, namely, that
neurons share the information at random. It would be therefore
interesting to find an experimental overlap that differs
signifficantly from $I(1)/H\{s\}$. Or, as
suggested by Gawne and Richmond, that the overlaps bear a spatial
dependence. That is to say, that even though the overall mean
redundancy might be close to $I(1) / H\{s\}$, neurons sitting
close to one another could be more redundant than pairs
standing further apart.

We would finally like to stress that the saturation of $I(N)$ is
only determined by the set of stimuli. One should not deduce that
when reaching the ceiling, the pool of neurons has reached its
maximum representational capacity. Rather, the stimuli are
no longer adequate to explore the coding capabilities of the
cells. If an experimentalist is interested in measuring
the representational capacity of the system, he or she should choose
a set of stimuli whose entropy is large enough so that he can determine
precisely the slope of the initial linear rise, that is, $I(1)$. If
the aim is to characterize how redundant the messages conveyed by
different neurons are, then a measure of the departure
from the linear rise is needed. Therefore,  $H\{s\}$ should not be
extremely large, as compared to $I(1)$.  We suggest that a
comparison between the measured value of $y$ with $I(1)/H\{s\}$ can
indicate if the ceiling effects are only because the set of stimuli is
finite, or whether they arise from some special organization in the
way different neurons represent information.

\ack
I thank Alessandro Treves and Dami\'an Zanette for very
useful discussions.
This work has been supported with a grant of
the Human Frontier Science Programm, number RG 01101998B.

\References

\item[] Abbot L F and Dayan P 1999 {\it Neural Comp.} {\bf 11}
91 - 101

\item[] Abramowitz M and Stegun I A 1972
{\it Handbook of mathematical functions} (Dover: New York)

\item[] Gawne T J and Richmond B J 1993 {\it J.
Neurosc.} {\bf 13} (7) 2758 - 2771

\item[] Karbowski J (2000) {\it Phys. Rev. E} {\bf 61} 4235
                                                                                   E 61, 4235 (2000)

\item[] Oram M W, Foldiak P, Perett D I and Sengpiel F 1998,
{\it Trends in Neurosc.} {\bf 21} 259 - 265

\item[] Panzeri S Treves A Schultz S and Rolls E 1999a
{\it Neural Comp.} {\bf 11} 1553 - 1577

\item[] Panzeri S, Schulz S R, Treves A, Rolls E T 1999b
{\it Proc. R. Soc. Lond B} {\bf 226} 1001 - 1012

\item[] Rolls E T Treves A and Tovee M J 1997
{\it Exp. Brain Res} {\bf 114} 149 - 162

\item[] Samengo I and Treves A 2000, accepted in {\it Phys.
Rev. E}, scheduled for 01 January 2001.

\item[] Shannon C E 1948
{\it AT \& T Bell Labs. Tech. J.} {\bf 27} 379 - 423

\endrefs

\end{document}